\documentclass[aps,prb,amsmath,twocolumn,amssymb,titlepage,superscriptaddress]{revtex4-1}
\usepackage{setspace}
\usepackage{float}
\usepackage{graphicx}
\usepackage{nicefrac}
\usepackage{amsfonts}

\usepackage{amssymb}
\usepackage{amsmath} 

\usepackage{subfigure}
\usepackage{multirow} 
\usepackage{tabularx} 
\usepackage{array}

\usepackage{units}

\usepackage{tensor} 
\usepackage{braket}

\usepackage{bm}
\usepackage{hyperref}
\usepackage{color}

\begin{document}
\title{Symmetry broken spin reorientation transition in epitaxial MgO/Fe/MgO layers with competing anisotropies}

\author{Isidoro Mart\'inez}
\affiliation{Dpto. Fisica de la Materia Condensada, IFIMAC and INC, Universidad Autonoma de Madrid, 28049, Madrid, Spain}
\author{Coriolan Tiusan}
\affiliation{Center of Superconductivity, Spintronics and Surface Science (C4S), Technical University of Cluj-Napoca, 400114 Romania}
\author{Michel Hehn}
\affiliation{Institut Jean Lamour, Nancy-Universit\'e Vandoeuvre Les Nancy Cedex, 54506 France}
\author{Mairbek Chshiev}
\affiliation{Universit\'e Grenoble Alpes, CEA, CNRS, INAC-SPINTEC, 38000, Grenoble, France}
\author{Farkhad G. Aliev}\email[Email: ]{farkhad.aliev@uam.es}
\affiliation{Dpto. Fisica de la Materia Condensada, IFIMAC and INC, Universidad Autonoma de Madrid, 28049, Madrid, Spain}
% Repeat:\maketitle
% \author{}
% \affiliation{}
% For each author

\begin{abstract}

The observation of perpendicular magnetic anisotropy (PMA) at MgO/Fe interfaces boosted the development of spintronic devices based on ultrathin ferromagnetic layers. Yet, magnetization reversal in the standard magnetic tunnel junctions (MTJs) with competing PMA and in-plane anisotropies remains unclear. Here we report on the field induced nonvolatile broken symmetry magnetization reorientation transition from the in-plane to the perpendicular (out of plane) state at temperatures below 50K. The samples were 10 nm thick Fe in MgO/Fe(100)/MgO as stacking components of V/MgO/Fe/MgO/Fe/Co double barrier MTJs. Micromagnetic simulations with PMA and different second order anisotropies at the opposite Fe/MgO interfaces qualitatively reproduce the observed broken symmetry spin reorientation transition. Our findings open the possibilities to develop multistate epitaxial spintronics based on competing magnetic anisotropies.

\end{abstract}
\maketitle

\section{Introduction}
%\begin{equation*}
%K_{eff}=K_v+K_{s}^{(I)}/t+K_{s}^{(II)}/t
%\end{equation*}

Magnetic films with the magnetization oriented perpendicular to the film plane are currently the best candidates for magnetic storage devices with respect to the challenge to decrease the bit size. The phenomenon of having a preferential magnetization perpendicular to film plane is usually referred to as Perpendicular Magnetic Anisotropy (PMA) \cite{Johnson1996ReviewAnisotropies,Dieny2017}. The bottleneck to have PMA is the control of the magnetic anisotropy which is characterized by an effective anisotropy constant ($K_{eff}$) that has a volume contribution  $K_v$ and two surface or interface contributions $K_s$ \cite{Bruno,Bruno1988,Dieny2017}. As a result, it can be described as $K_{eff}=K_v+K_{s}^{(I)}/t+K_{s}^{(II)}/t$ where $K_{s}^{(I)}$ and $K_{s}^{(II)}$ are the surface anisotropies at the lower (I) and upper (II) interfaces and $t$ is the ferromagnetic layer thickness. The volume contributions arise from magnetocrystalline anisotropy, magnetoelastic anisotropy and shape anisotropy. The latter contribution induces commonly an in-plane magnetized configuration in magnetic thin films. However, in ultrathin layers several angstroms thick, the surface contribution to anisotropy can exceed the volume shape anisotropy contribution leading to the PMA. Surface contributions are linked to roughness and interface alloy, strain and mainly to the broken symmetry at the interface or at the surface of the magnetic layers. It manifests itself by the appearance of a second order anisotropy \cite{Dieny2017}. The constant $K_{s1}$  of the first order surface anisotropy energy per unit area may range between $K_{s1} \approx 1 \times 10^{-3} \, \text{J/m}^2$ in ultrathin Co,Fe and Ni films \cite{Bruno1988} and $K_{s1} \approx 3-4 \times 10^{-3} \, \text{J/m}^2$ at CoFeB/Pt interfaces \cite{Ngo2014}.

The recent trends in spintronics using MTJs also take advantage of PMA to provide large tunneling magnetoresistance (TMR), enhanced thermal stability \cite{naturemat}, low spin torque switching currents \cite{Leutenantsmeyer2015,Lau2016} and record small lateral sizes \cite{Igarashi2017}. Those features are critical for the progress towards spin transfer torque based magnetic random access memories. It appears that MgO/Fe interfaces show PMA substantially exceeding the values reported for the prototype Co/Ni(111) system \cite{Gottwald2012}.

The spin-orbit interaction (SOI) emerging from the reduced interfacial symmetry of the Fe \textit{d}-orbitals and O \textit{p}-orbitals has been suggested as a main source of PMA at MgO/Fe interfaces \cite{Yang2011,Hallal2013}. First-principles calculations give $K_{s1}$ ranging between $1.5 \times 10^{-3} \, \text{J/m}^2$ and $1.8 \times 10^{-3} \, \text{J/m}^2$ \cite{Hallal2013} in a reasonable agreement with experiments \cite{naturemat,Koo2013,Lambert2013}. According to its interfacial nature, the PMA intensity varies strongly with the magnetic layer thickness. This means a great challenge to increase the critical thickness of the spin reorientation transition \cite{naturemat} above a few nm, without the need of the permanent application of an external magnetic field stimulus \cite{Kozioł-Rachwał2013}. By changing the normal metal (NM) in NM/Fe/MgO (NM=V; Cr) \cite{Lambert2013}, the transition between out of plane and in-plane anisotropy remains between 4 and 6 Fe monolayers. Reducing the bulk magnetization through bulk Fe doping by V or Cr impurities reduces the easy-plane demagnetizing energy and slightly changes the critical thickness at which reorientation occurs  \cite{Hallal2014}. Finally, decreasing the temperature down to 5 K, leads to the PMA and saturated magnetization enhancement. This has a rather limited impact on the critical thickness for the spontaneous out of plane magnetization alignment \cite{Gabor2015,Fu2016}.

\begin{figure}[h]
\centering
	\includegraphics[width=0.75\linewidth]{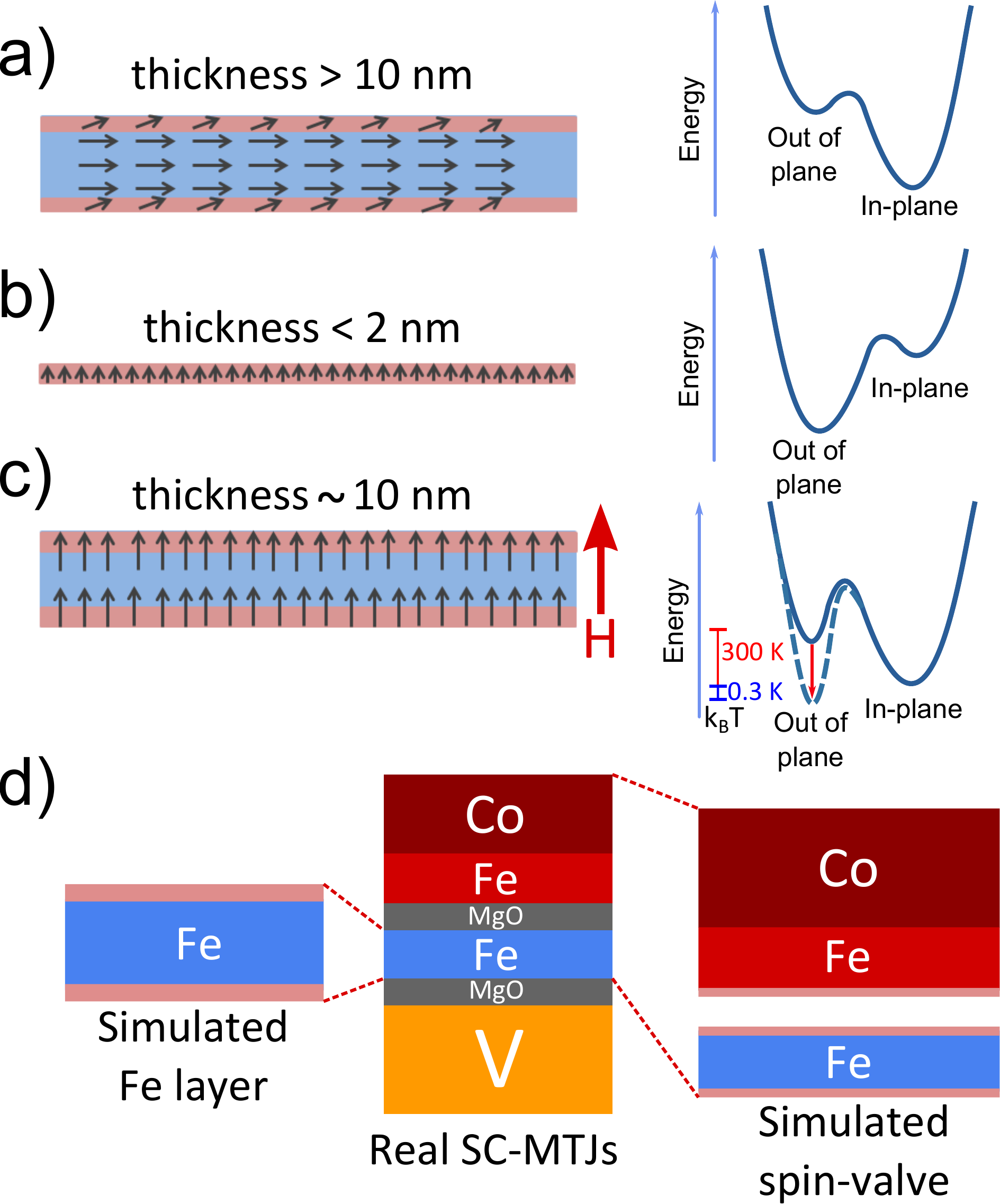}
	\caption{Sketch explaining the creation of non-volatile out of plane magnetization in thin ferromagnetic films showing the competing PMA and shape anisotropies. (a) For relatively thick films, due to shape anisotropy, the ground state is in-plane magnetized. (b) In thin enough films (typically $<$2 nm), due to the dominant PMA,  the ground state is magnetized out of plane already at room temperature. (c) Case of competing PMA and shape anisotropies taking place in intermediate thick films (10 nm here). If an out of plane magnetic field is applied and removed at sufficiently low temperatures (indicated by vertical colour bars), the magnetization can switch from in-plane to the non-volatile out of plane state. The lower part (d) sketches the samples under study (middle) and the design of the simulated structures (left$\rightarrow$ soft MgO/Fe/MgO structure and right$\rightarrow$ soft layer weakly coupled to the hard layer, see text for details).}
	\label{esquema}
\end{figure}

Seminal MTJs grown by Molecular Beam Epitaxy (MBE) have in-plane anisotropy and typically incorporate about 10 nm thick Fe or FeCo soft ferromagnetic layers (FM) separated by MgO barriers from the magnetically hard layers \cite{Tiusan2007,Guerrero2007}. According to our discussion, due to the shape anisotropy, such soft FM layers remain in-plane magnetized at room temperature and zero magnetic field. However, a ferromagnetic resonance analysis pointed out the presence of a PMA contribution to the total anisotropy \cite{Belmeguenai2013} even for in-plane magnetized FM electrodes. As a result sketched in Figure \ref{esquema}a, the potential profile exhibits an energy minimum with in-plane magnetization but presents a metastable out of plane magnetization state with a local energy minimum. Decreasing the ferromagnetic layer thickness will enhance the influence of PMA \cite{Nistor2010}. Therefore, for a film thickness below some critical value (typically around 1-2 nm) the out of plane magnetic state becomes the ground state as sketched in Figure \ref{esquema}b. 

The main concept of the ferromagnetic system with competing in-plane and out of plane anisotropies is sketched in Figure \ref{esquema}c. By combining low temperatures with a brief application of a perpendicular field, we will show that a non-volatile magnetization reorientation, as explained in Figure \ref{esquema}c, could be observed in MgO/Fe/MgO films with Fe(100) thickness the exceeding critical thickness of in-plane reorientation (here 10 nm). Once the external magnetic field is removed, the magnetization could remain perpendicular at low temperatures even in relatively thick films if the energy barrier between the metastable (perpendicular out of plane) and ground (in-plane) states substantially exceeds the thermal energy. This state emerges due to competing in-plane and perpendicular interfacial anisotropies sketched in Figure \ref{esquema}c. The main experimental observations are qualitatively supported by simulations. The breaking of the perpendicular magnetization symmetry has been explained as a consequence of the difference in disorder between the bottom and the top interfaces in MgO/Fe/MgO. To our best knowledge, this is the first report on magnetization orientation manipulation of this kind in magnetic tunnel junctions.

\section{Experimental results}

\subsection{Tunneling magnetoresistance}

\begin{figure}[h]
\centering
\includegraphics[width=0.7\linewidth]{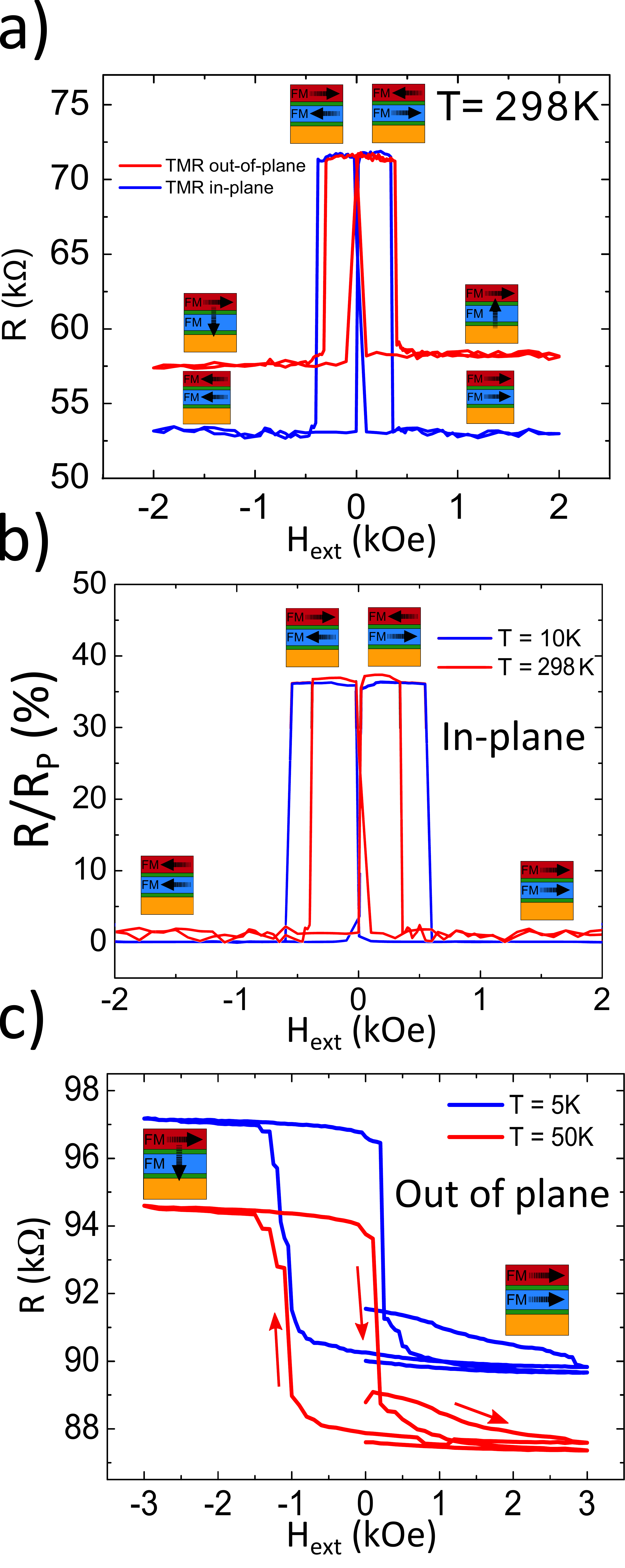}
\caption{(a) In-plane and out of plane resistance vs. field curves at room temperature. Part (b) shows in-plane resistance (normalized by P state) vs. field curves for different temperatures. (c) Out of plane resistance vs. field curves at different temperatures. The coercive fields of the FM electrodes are higher in the out of plane direction ($H_{C-soft}\simeq1\text{ kOe}$ and $H_{C-hard}>5\text{ kOe}$). Red arrows indicate the magnetic field sweep history.}
\label{TMRsRT}
\end{figure}

Sample growth and measurement techniques are explained in the Methods section. Figure \ref{TMRsRT}b shows in-plane and out of plane TMR at different temperatures. At room temperature the in-plane TMR is close to 36 \%. Instead the out of plane TMR is lower (around 24 \%) because in that case, as indicated in Figure \ref{TMRsRT}a, the two electrodes are perpendicularly oriented for the maximum resistance values. Both in-plane and out of plane TMR values are somewhat lower than those typically reported for single barrier MTJs. The cause is that the device under study is composed of two tunnel junctions: a normal tunnel junction between Fe and V connected in series with a MTJ (formed by Fe and Fe/Co). The first (V/MgO/Fe(10nm)) junction is expected to have a tunneling resistance independent of the FM orientation above the critical temperature of V (if the contribution of anisotropic tunneling magnetoresistance is neglected). The main function of the normal metal (Vanadium) electrode is as single crystalline (001) oriented buffer template and contact the 10 nm Fe layer under investigation through a thin MgO barrier. Vanadium is one of the few nonmagnetic and conducting materials, which can be epitaxially grown in single crystal Fe/MgO(001) tunnel junctions. Vanadium has a bcc structure perfectly compatible with the epitaxial subsequent growth of MgO and Fe with (001) orientation. The inspected soft Fe layer is then contacted through a second MgO barrier by the Fe/Co hard layer which plays the role of a spin-orientation sensor. The room temperature thermal energy helps to overcome the energy barrier needed to flip the magnetization direction. In the conditions of competing anisotropies at room temperature this leads to similar switching fields from upwards to downwards effect to that between opposite in-plane directions as reflected in Figure \ref{TMRsRT}a. The thermal effects are also reflected in Figure \ref{TMRsRT}b, where the coercive fields of the in-plane oriented electrodes clearly decrease with increasing temperature.

Our main experimental finding has been observed at low temperatures below 50K is represented in Figure \ref{TMRsRT}c. We apply at low temperatures, an out of plane magnetic field not exceeding 3 kOe to maintain the in-plane magnetization orientation of the hard electrode unchanged. We observed then a change in resistance corresponding to a magnetization flip of the soft Fe electrode from the in-plane ($H_{ext}=0$) to the out of plane direction at about $H_{ext}=-1$ kOe. In the range of the explored perpendicular magnetic fields ($<$3 kOe) and low temperatures (T$<$50 K) this magnetization switch (from in-plane to out of plane direction and backwards) takes place only in the negative field direction as shown in Figure \ref{TMRsRT}c. Our claim on the magnetization switch asymmetry is based on the following experimental observation. It shows that for sufficiently low temperatures, at least below 50K, for the the positive field direction, the soft electrode retains its direction close to in-plane. Such an asymmetric magnetization response is relatively robust to the history of the variation of the magnetic field (see Supplemental Figure 1) and disappears at room temperature  (Figure \ref{TMRsRT}a). The possible sources of the magnetization flip asymmetry will be discussed further below.

\begin{figure}
	\centering
	\includegraphics[width=0.85\linewidth]{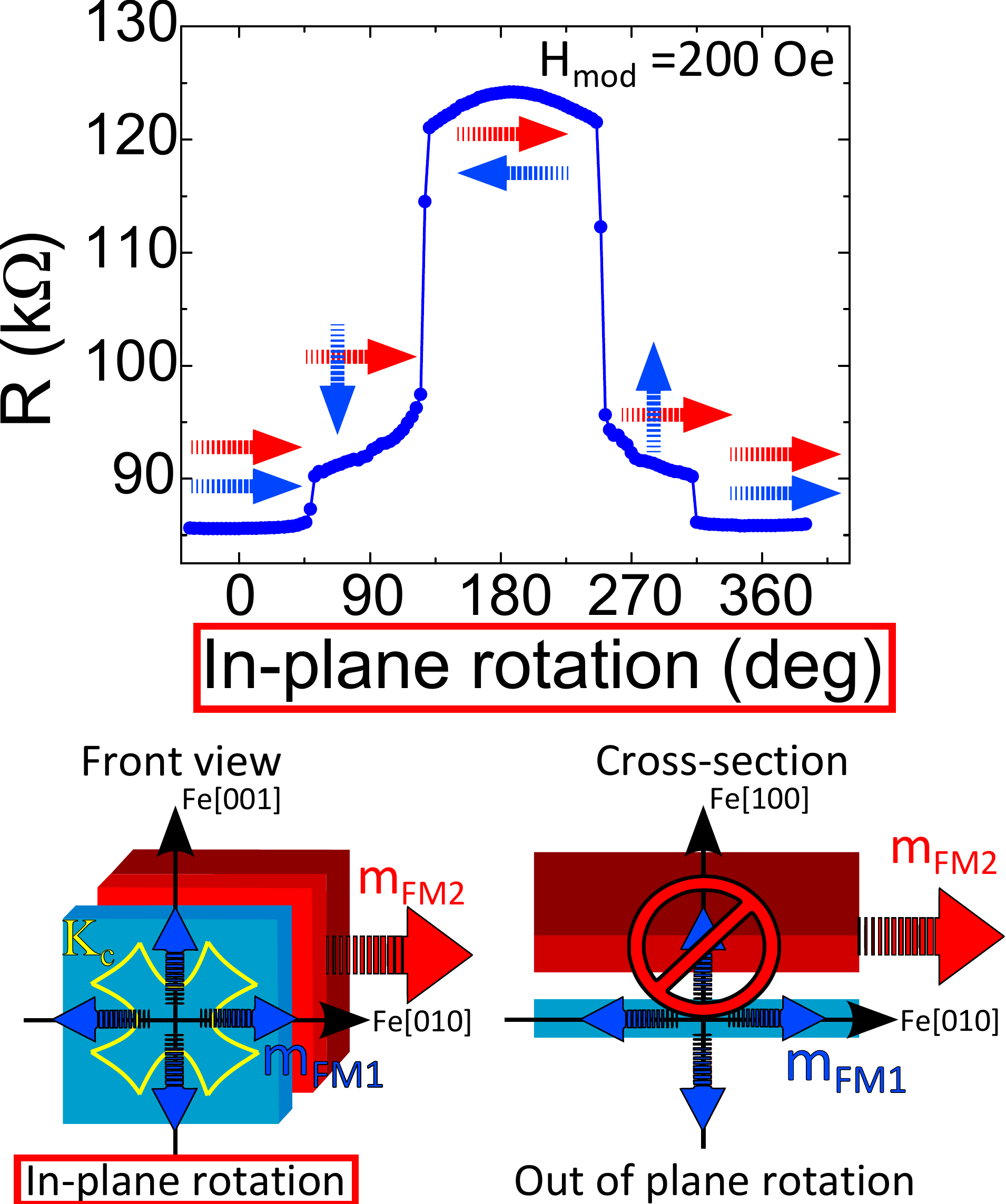}
	\caption{The top part shows the resistance vs. field curve (measured at T=5K) of an in-plane rotation of the external magnetic field with modulus $H_{mod}=200$ Oe. This field value maintains the hard layer practically fixed and only rotates the soft Fe layer. One observes a nearly symmetric response (sketched in the bottom left part). This excludes a possible explanation of the observed in-plane to out of plane reorientation transition (sketched in the bottom right part) in terms of the in-plane response due to magnetic field misalignment. The FM1 magnetization is depicted by blue arrows (Fe soft layer) while the FM2 magnetization is depicted by red arrows (Fe/Co hard layer).}
	\label{fig:In_plane_Rotation_TMR_2}
\end{figure}

For the experiments with out of plain field one unavoidably has some small (in our case $\leq3$ degrees) deviation of the applied field from a strictly perpendicular direction. We cannot therefore exclude an in-plane component of the external field to be present during such experiments. We have tried to verify a possible explanation of the transition as due to in-plane magnetization rotation induced by the above mentioned misalignment (between the normal vector of the interfacial plane and the external field). With this aim we have studied the tunneling resistance under a full in-plane magnetic field rotation of the soft layer magnetization keeping the hard layer fixed. The experimental results (Figure \ref{fig:In_plane_Rotation_TMR_2}) point out towards a 4-fold symmetry of the soft Fe layer magnetization rotation. This is expected for the epitaxial soft FM layer in MgO/Fe(001)/MgO. The bottom left part in Figure \ref{fig:In_plane_Rotation_TMR_2} summarizes through a sketch the presence of the 4-fold symmetric response of the soft layer when the external field is rotated within the interface (in-plane). We insist therefore on the strictly out of plane nature of the observed asymmetric spin reorientation transition as sketched at the bottom right part of Figure \ref{fig:In_plane_Rotation_TMR_2}.

The resistance values in three well defined magnetic states of the soft layer allow an evaluation of the effective spin polarization of the FM electrodes at corresponding temperatures. Following a simplified model \cite{conductance} we approximate the conductance of a whole structure formed by a tunnel junction (TJ) and a MTJ in series through the following expression:  $G^{-1}=G_1^{-1}+[G_2 (1+p^{2} cos \vartheta)]^{-1}$. Here $G_1$ is the low bias conductance of the V/MgO/Fe junction, $G_2$  is the low bias conductance of the MTJs in the perpendicular state, $p$ is the effective spin polarization and $\vartheta$ the angle between the magnetizations of the ferromagnets. From the room temperature out of plane and in-plane resistance vs. field measurements shown in Figure \ref{TMRsRT} we obtain three different conductance values corresponding to the three magnetic states: parallel ($\vartheta=0$), anti-parallel ($\vartheta=\pi$) and perpendicular ($\vartheta=\pi/2$) states. The knowledge of those conductances provides $p=0.65$ at 300K and $p=0.68$ at 10K. 

The estimated decrease of spin polarization with increasing temperature could be attributed to thermally excited magnons. We also note that the obtained $p$ values underestimate the effective spin polarization because they were obtained without consideration of the additional shunting effect coming from the coherent tunneling between V and top Fe across the two MgO barriers. This effect is particularly important for majority spins whose coherence length in Fe is known to be larger than 10 nm. The above mentioned effect enhances the AP conductivity and reduces the expected large TMR. Moreover, we mention that the evaluated $G_{1,2}$ conductances increase with temperature proving high quality and pinhole free barriers.

On the other hand, we find the second MgO barrier (between the 10 nm Fe and the Fe/Co hard electrode) to be about 4 times more transparent than the bottom MgO layer (i.e. $G_{2} \simeq 4G_{1}$). This points towards an accumulation of the structural disorder during the epitaxial stack growth and could explain the negligible variation of TMR of the whole structure including two MgO barriers with different structural qualities. Note that Ref. \cite{Cascales2012} also observed indirectly an accumulated disorder in the epitaxial double barrier magnetic tunnel junctions through a substantial difference in the two MgO barrier transparencies with the same nominal thickness.

\section{Simulations}

\begin{figure}
	\centering
	\includegraphics[width=0.9\linewidth]{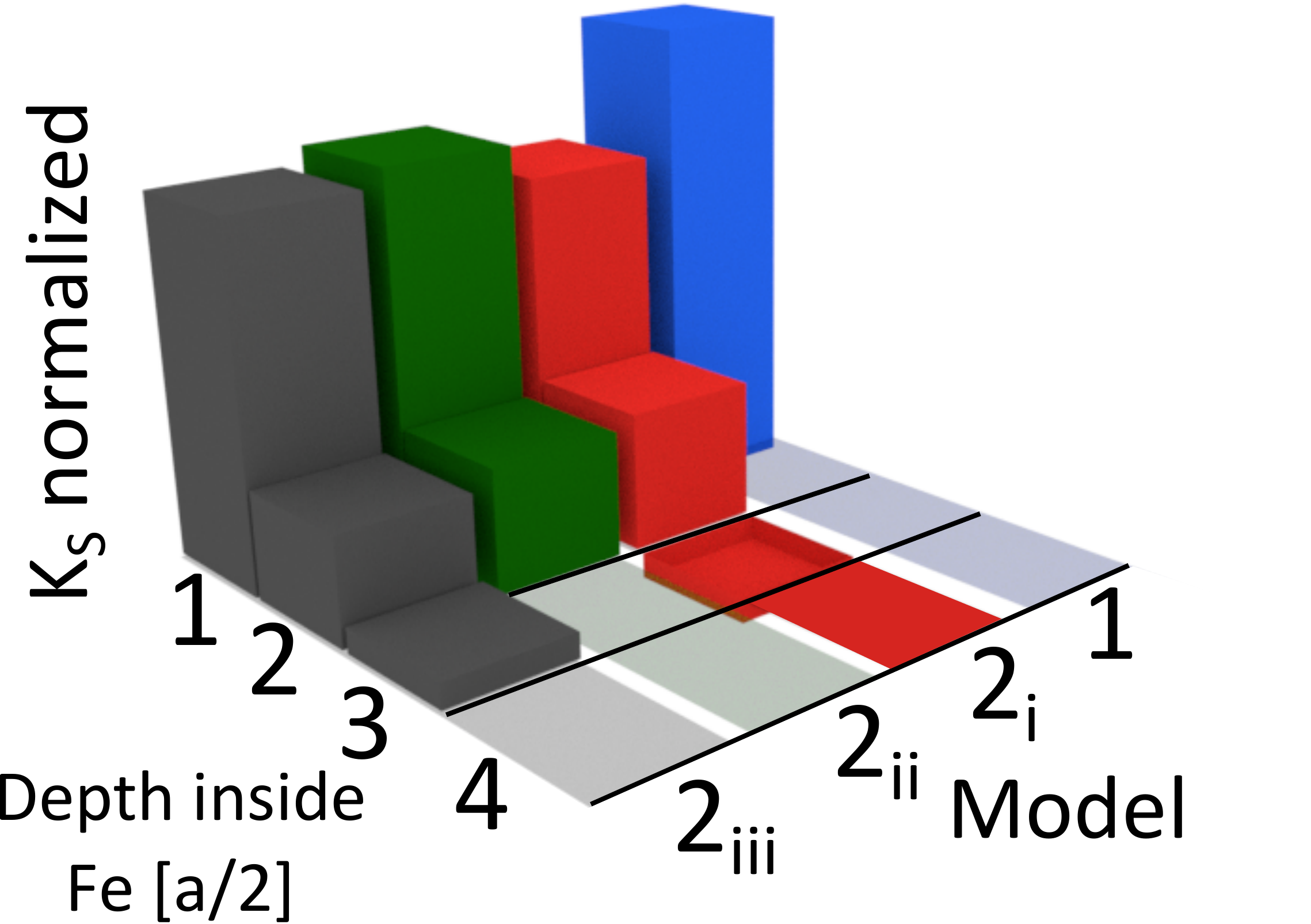}
	\caption{Sketch of the different possible distributions of the surface anisotropy $K_{s1}$ explored in simulations. Blue prism represents PMA variation within model M1. Three possibilities within models M2(i-iii) explored are represented correspondingly by red, green and grey coloured prisms respectively. The total volume energy corresponding to PMA has been kept constant.}
	\label{fig:repartomodelos}
\end{figure}

The details on simulations are explained in the Methods section. We first discuss the models used to simulate the magnetization reversal of the symmetric soft Fe layer (without $K_{s2}$ anisotropy) with competing anisotropies under applied perpendicular magnetic field.  A perpendicular magnetization anisotropy has been introduced using two qualitatively different ways sketched in Figure \ref{fig:repartomodelos}. The first approach (model $M1$) uses a single step PMA variation within the first atomic layer. The concentration of $K_{s1}$ in the first layer only corresponds to the vacuum/Fe/vacuum case discussed in Ref.\cite{Hallal2013}. The second (model $M2$) approach involves three slightly different versions labeled ($M2\, i-iii)$ which allow the different PMA variation in steps of a half lattice period $a/2$ (an Fe atomic layer). The model ($M2\, i)$ with an oscillatory decay of PMA inside Fe is the closest to the numerical predictions for the MgO/Fe/MgO case\cite{Hallal2013}.

The model $M2i$ uses the following uniaxial surface anisotropy $K_{s1}$ distribution in percentages: 65\% for the first layer of Fe atoms ($K_{s1}$) and 30\% for the second one. The third layer, following DFT results \cite{Hallal2013}, is assigned an uniaxial in-plane anisotropy of roughly 10\% in volume energy of the first layer of Fe atoms. Finally 5\% of $K_{s1}$ is assigned to the 4th layer. Generally, those percentages approximately follow the DFT results \cite{Hallal2013} concentrating $K_{s1}$ within the first four atomic Fe layers and approximating the surface anisotropy variation with the predicted Friedel-like decay oscillating between PMA and in-plane magnetic anisotropy (IMA). Other versions of the $M2$ model ($M2ii$ and $M2iii$) modify the Friedel-type PMA decay towards a more monotonous PMA variation. Below we describe our main experimental findings and compare the results with simulations within the above stated approaches $M1$ and $M2$.

\subsection{Simulation results on a single symmetric Fe layer}

Using the above introduced models we have carried out zero temperature micromagnetic simulations of the MgO/Fe/MgO structure shown in Figure \ref{esquema}d (bottom left) with the optional possibility of coupling between the soft and hard electrodes (bottom right). This paragraph mainly describes the most important simulation results related with the spin reorientation transition in a symmetric MgO/Fe/MgO structure with competing anisotropies. We shall discuss separately the possible influence of the accumulated structural disorder on the spin reorientation transition through the introduction of a higher order surface anisotropy term.

\begin{figure}
	\centering
	\includegraphics[width=1.0\linewidth]{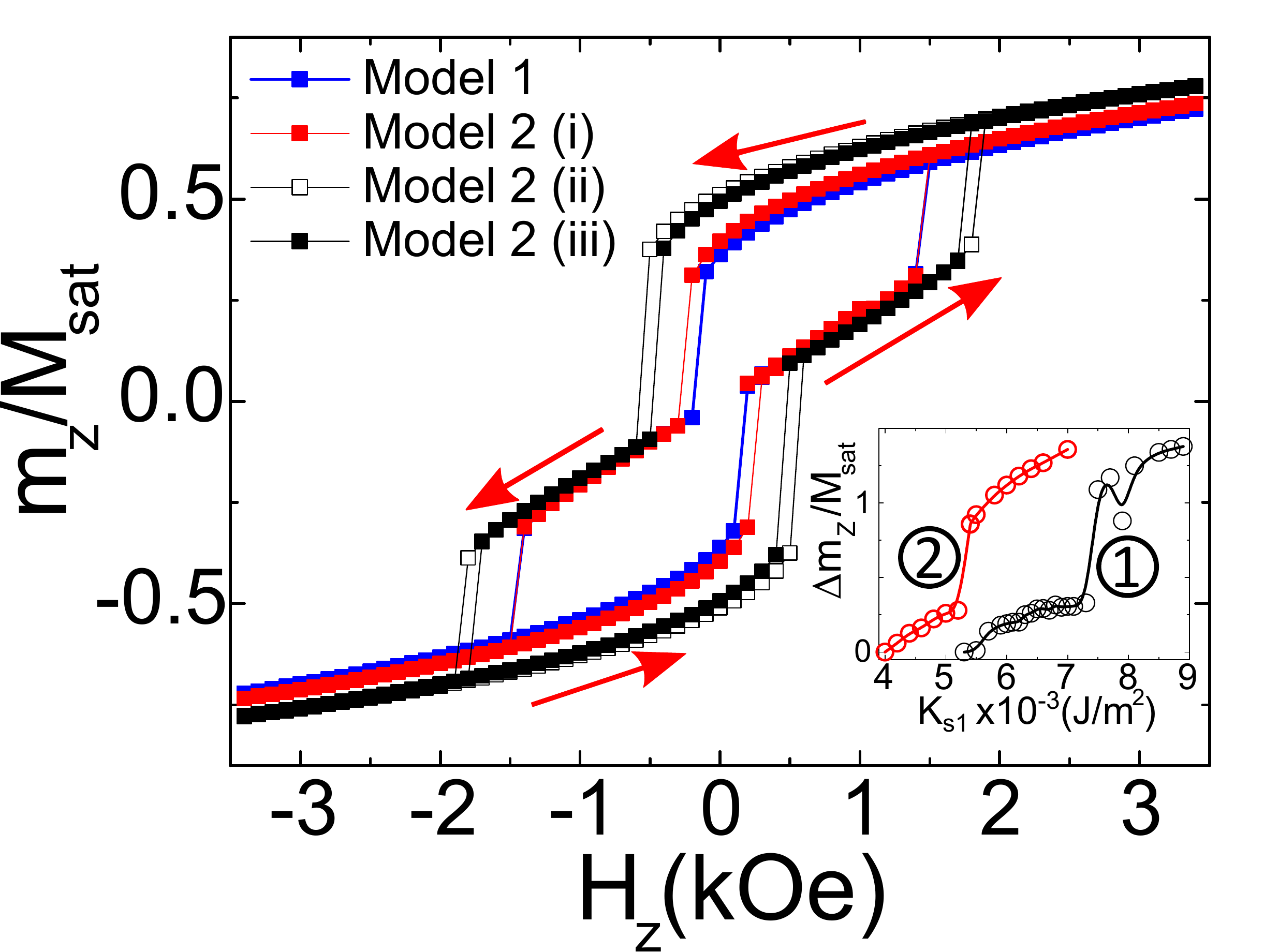}
	\caption{Simulation of the normalized (by $M_{sat}$) perpendicular to the interface $z$ component of the magnetization of the interfacial layer of the 10 nm thick Fe layer. In MgO/Fe/MgO a PMA of $6.35\times10^{-3}\,\text{J/m}^2$ has been used. The inset represents the variation of the maximum normalized magnetization jump during the spin reorientation transition on $K_{s1}$ when simulated within the model $M2i$ for two different saturation magnetizations. This inset provides estimation of the anisotropy needed to create the competing anisotropy conditions in MgO/Fe(10nm)/MgO. Curve 1 uses a constant value of $M_{sat}=1700 \times 10^{3} \, A/m$ while curve 2 uses a $25\%$ reduced interfacial values of $M_{sat}$. The red arrows indicate the magnetic field sweep history.}
	\label{fig:ModelosSingle}
\end{figure}

Figure \ref{fig:ModelosSingle} compares the simulation results obtained with models $M1$ and $M2$ involving three different PMA energy distributions keeping the total PMA energy fixed. The introduction of the Friedel-like PMA variation \cite{Hallal2013} (model $M2i$) softens the reorientation transition with respect to the $M2ii-iii$ models. Although the $M1$ PMA distribution also softens the reorientation transition, it reduces the non-volatility of the zero field magnetic state, in contrast to the experiment.

The inset to Figure \ref{fig:ModelosSingle} shows that the magnetization flip does not happen for the PMA values below approximately $K_{s1}=5 \, \times10^{-3}\,\text{J/m}^2$. On the one side, to reproduce the observed magnetization flip between the near in-plane and the near out of plane magnetizations, modeling has to use $K_{s1}$ values not exceeding approximately $6.4 \, \times10^{-3}\,\text{J/m}^2$. These PMA values are somewhat bigger than those provided by DFT\cite{Hallal2013} calculations. For the surface anisotropy exceeding $6.4 \, \times10^{-3}\,\text{J/m}^2$ a reorientation transition tends to take place directly between two nearly perpendicular magnetization states. This means without locking the magnetization in the intermediate, close to in-plane state. For confirmation see inset of Figure \ref{fig:ModelosSingle} showing a second upturn in the magnetization jump. The supplemental Figure 2 shows that the spin flip transition is rather weakly affected by the presence of the Fe layer cubic anisotropy. This demonstrates that it is mainly a demagnetization energy contribution which competes with PMA to provide the spin reorientation transition. 

Our experimental results point towards a relatively weak coupling between the MgO/Fe/MgO layer and the hard Fe/Co electrode. Indeed, simulations show that the in-plane to out of plane spin reorientation transition could be affected by either ferromagnetic or antiferromagnetic coupling between free and sensing electrodes (see  Supplemental Figure 3).

\section{Discussion}

The observed low temperature non-volatile perpendicular magnetization state in 10 nm thick Fe(100) layers could be a specific feature of ferromagnetic films with competing PMA and IMA. That is to say with thickness few times exceeding the critical one where a perpendicular magnetic state becomes the ground state. We used a hard ferromagnetic layer to sense the effect. Nevertheless one may anticipate that a reorientation transition could be also detected by measuring the anisotropic magnetoresistance at sufficiently low temperatures through the application of magnetic fields along two different (in-plane or out of plane) directions and with the current directed within the film plane.

The observed reorientation transition shows two remarkable properties at low temperature: non-volatility and an asymmetric magnetization response to the out of plane external field. Simulations taking into account PMA have reproduced the experimentally observed non-volatility. However the presence of the magnetization asymmetry was not reproduced in the case of symmetric MgO/Fe/MgO structures. The magnetization asymmetry points towards some net field acting perpendicularly to the interface direction. This breaking symmetry could originate from differences in the crystalline disorder at the two MgO/Fe interfaces and will be discussed separately below.

\subsection{Sources of discrepancy between experiment and simulations}

Despite the general qualitative agreement between experiment and simulations, a certain number of open questions remain yet to be answered. Our simulations show that the magnetization flip occurs at fields around 1- 2 kOe with the magnetization after the flip being not fully perpendicular to the interface (Figure \ref{fig:ModelosSingle}). The switching field is nearly independent on the upper saturation field when varied in the range 3.5-7 kOe (not shown). On the other side, the experiments reveal a near in-plane to near out of plane spin flip at about or below 1 kOe (Figure \ref{TMRsRT}c). A different issue is the reported PMA value needed to reproduce the abrupt magnetization switch in the perpendicular magnetic field. This value is somewhat larger than previously reported.

Several factors, not present in the simulations, could be responsible for such differences. Among them are the (i) presence of defects, (ii) a modified interfacial saturation magnetization and/or (iii) the presence of finite temperature in the experiments, among others. Some reduction of the lateral size of the simulated structure affects competing anisotropies leading to the suppression of the robustness in the simulation results. Our discussion is centered mainly on physical effects and omits the possible influence of chemical bonding on PMA\cite{Mlynczak2013}, because the latest numerical studies minimize the effect of  Fe-O \textit{p}-\textit{d} hybridization on the PMA\cite{Odkhuu2016}.

Let us focus on the possible influence of the electrodes coupling on the results. If the hard FM layer was magnetostatically and/or weakly exchange coupled to the soft Fe layer under study, this would give rise to a more complex magnetization reversal (see  Supplemental Figure 3) with other possible states \cite{Ummelen2017} well beyond those simple three relative magnetization states observed and discussed above for the uncoupled MgO/Fe/MgO.

The presence of asymmetrically located interfacial defects could lead to higher order contributions to the PMA \cite{Timopheev2016}. Among other possible sources of the discrepancies could be the value of the effective Fe moment in the proximity to the Fe/MgO interface. Some reports point towards interfacial magnetic moments at ferromagnet/oxide interfaces enhanced up to 25\%\cite{Jal2015,Ueno2016}. Our simulations, however, show that the critical values of the surface anisotropy (i.e. $K_{s1}$ values) needed for the spin reorientation transition approach towards those obtained by DFT, once we reduced the saturation magnetization at both interfaces (see insert in Figure \ref{fig:ModelosSingle}).

An additional discrepancy in the PMA values could be the lattice mismatch (stress) providing interfacial electric fields \cite{Hallal2013} and stronger pinning of interfacial magnetic moments\cite{Koziol2013}. Just as small as a 0.5 $\%$ reduction of the lattice parameter is expected to increase $K_{s1}$ from $(1-2) \, \times10^{-3}\,\text{J/m}^2$ to $(5-6) \, \times10^{-3}\,\text{J/m}^2$ at the MgO/FeCo interface. Moreover, the PMA for the individual Fe atoms deposited on MgO(100) thin films could increase $K_{s1}$ in the same order of magnitude \cite{Baumann2015}. So far the PMA has been investigated in about 1-2 nm thick ferromagnetic layers because only ultrathin magnetically soft layers provided conditions for the room temperature operation of MTJs with PMA \cite{Meng2011}.

\subsection{Possible origin of the out of plane bias field}

\begin{figure}[h!]
\centering
\includegraphics[width=0.8\linewidth]{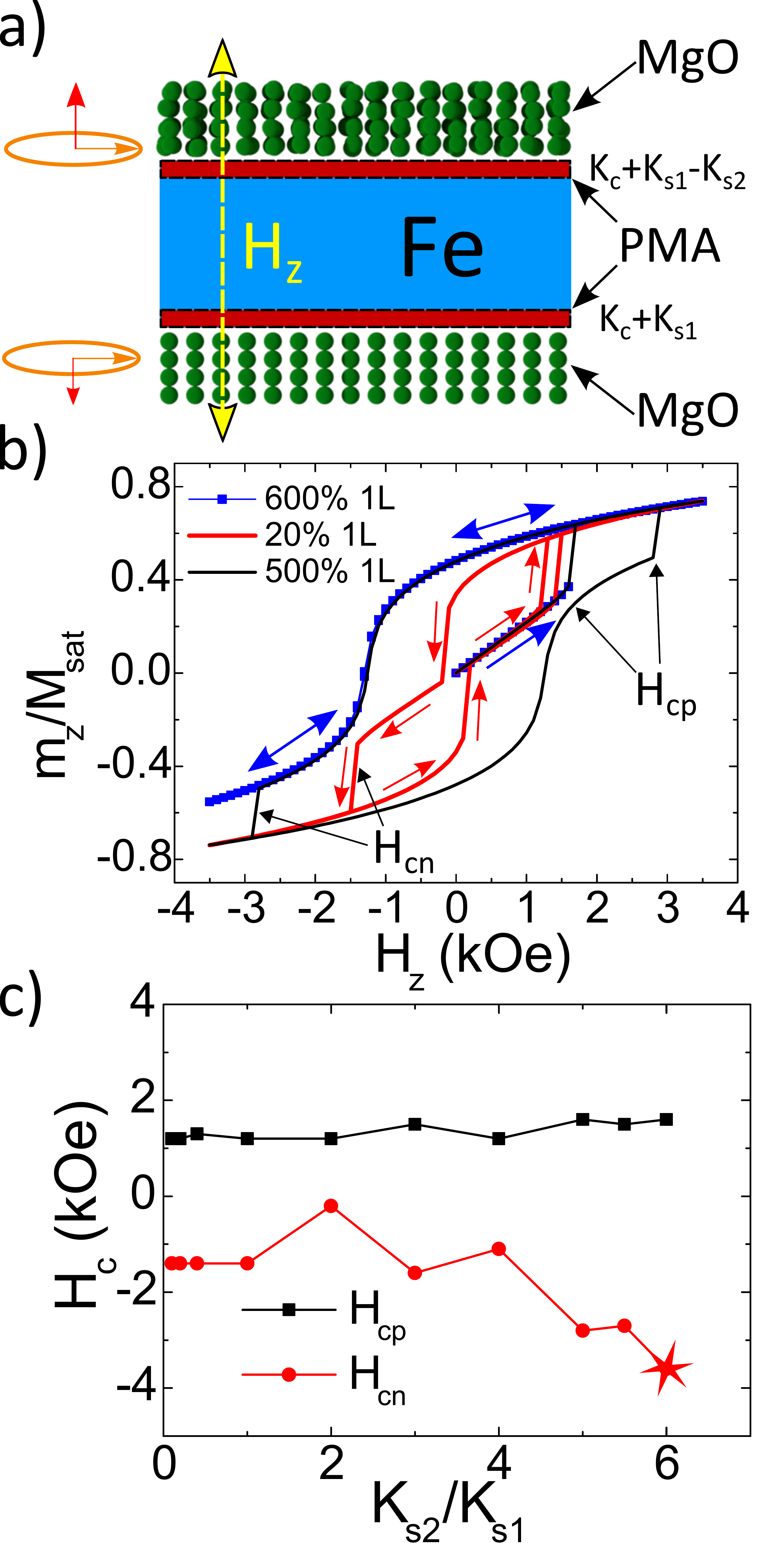}
\caption{The left side of part (a) explains the presence of the net Rashba field component perpendicular to the interface as a consequence of the difference in the upper vs. bottom longitudinal electric field components. This is due to the different degree of disorder at the interfaces. The right part in (a) explains an alternative possibility for the appearance of the second order surface anisotropy at the more disordered (top) interface. Part (b) shows two simulated magnetization loops (within model $M2i$) with different relations between the second order and the interfacial anisotropy terms. The coloured arrows (related to the blue and red cycles respectively) show the magnetic cycle step by step sweep. Part (c) shows how the asymmetry of the positive ($H_{cp}$) and negative ($H_{cn}$) coercive fields, corresponding to the transition between the near in-plane and near out of plane magnetizations, emerges with the increase of the relative contribution of the second order anisotropy at the Fe/MgO interface. The point marked as a star shows a negative coercive field out of the actual field range.}
\label{fig:Model_Simulations_K2}
\end{figure}

Below we discuss in more detail the possible origin of the observed perpendicular magnetization switching asymmetry. The two possible scenarios sketched in Figure \ref{fig:Model_Simulations_K2}a are based on the difference in the interfacial disorder between the bottom and the top Fe/MgO interfaces. From the one side, such structural disorder could introduce a component of the Rashba field perpendicular to the interface which should be different at two interfaces with different structural disorder. From the other side, we have already discussed that from the crystallographic point of view, the quality of the bottom and top MgO barrier should not be similar. The reason is that when growing MgO on bottom V(001), the V-(001)-MgO(001) lattice mismatch is $2\%$, smaller than the Fe(001)-MgO(001) mismatch ($4\%$) implicated when growing the top MgO barrier on middle Fe. As a consequence, the plastic relaxation limit thickness of MgO will be smaller when grown on Fe than on V \cite{Bonell2010}. Both barriers have nominal 2nm thickness, well below the relaxation limit, so dislocations clearly occur but the dislocation density in the top barrier, grown on Fe, will be larger. This has direct consequences on symmetry filtering, the defects promoting additional tunneling channels and therefore increasing the top barrier transparency.

As a consequence, a higher (e.g. second) order interfacial uniaxial anisotropy $K_{s2}$ could show up differently at two distinctly disordered interfaces. Already back in 1994  Dieny and Vedyayev showed analytically \cite{Dieny1994} that spatial fluctuations of the film thickness with $K_{s1}=Const.$, and period of the fluctuations lower than the exchange length of the ferromagnet may lead to a higher $K_{s2}\cos^4(\Theta)$ contribution to the PMA in addition to the $K_{s1}\cos^2(\Theta)$ term. Here  $\Theta$ is the angle between magnetization and perpendicular to the interface axis. Such possibility has been corroborated by a recent report\cite{Timopheev2016} showing that magnetization reversal could be substantially modified for opposite $K_{s1}$ and $K_{s2}$ signs.

Within these lines, we have carried out detailed simulations in order to verify the possible influence of the $K_{s2}$ surface anisotropy term on the magnetization reversal. Figure \ref{fig:Model_Simulations_K2}b describes the modified simulations carried out within Model 2(i) where the competing anisotropies $K_{s1}$ and $K_{s2}$ have opposite signs. The $K_{s2}$ anisotropy has been introduced only at the upper Fe/MgO interface, more disordered by the growth history. For the negative $K_{s2}$ values exceeding few times $K_{s1}$, we observed a strong asymmetry in the hysteresis cycle magnetization switch from the near in-plane to the near out of plane direction which resembles the experiment.

Figure \ref{fig:Model_Simulations_K2}c shows how the asymmetry in the positive (in-plane to out of plane) coercive field $H_{cp}$ and the second jump (negative in-plane to out of plane) coercive field $H_{cn}$ vary with increasing  $K_{s2}/K_{s1}$ ratio. The last (maximum negative) switching field value is taken as just exceeding the maximum negative applied field. This is because the in-plane/out of plane transition does not take place in that case (within the field range used) and magnetization just returns along the same trajectory, representing a strong asymmetry in field magnetization response. This is qualitatively similar to the one observed experimentally. Although the asymmetry in spin reorientation transition appears only for $K_{s2}/K_{s1}>5$, the critical condition for the relation between the corresponding energy contributions is around 0.2-0.4. This is because the reorientation transition takes place at angles about $\Theta \approx 1.37$ rad where the relation between angular dependent factors is $[\cos^4(\Theta)]\ll[\cos^2(\Theta)]$. We note that a more uniform distribution of the $K_{s2}$ along the four (instead of the single) interfacial cells does not alter substantially the simulation results.

One could also speculate on the above mentioned perpendicular component of the Rashba field induced by the oxygen defects inside the MgO and/or lattice mismatch at the Fe/MgO interface. Both factors could give rise to a local in-plane component of the electric field $E_{per}$ in addition to the well-established perpendicular interfacial electric field $E_{per}$. Since the Rashba field $B_R$ is proportional to the cross product of electron momentum $k$ and electric field $E$, and these depend on the local interface disorder, each of the Fe/MgO interfaces could generate locally different Rashba field components perpendicular to the interface $B_{Rz}$. Indeed, in the real growth conditions, the interfacial defects concentrations and their type should be different for the bottom (MgO/Fe) and the upper (Fe/MgO) interfaces. One therefore could envisage on average a difference in the perpendicular components of the corresponding Rashba fields as sketched in the left part of Figure \ref{fig:Model_Simulations_K2}a. The presence of such bias field could explain the existence of the field induced reorientation transition for one of the field directions only (at least within the field range under study).

\section{Conclusions}

An abrupt magnetic field induced transition between in-plane and out of plane magnetization states have been observed and investigated experimentally through magnetoresistance and by simulations in MgO/Fe/MgO layers with competing anisotropies. At sufficiently low temperatures (below 50K), the out of plane magnetization state becomes asymmetric in field and non-volatile, most probably due to the difference in disorder between the two Fe interfaces.
 The presence of three different remanent magnetic states in V/MgO/Fe/MgO/Fe/Co MTJs, potentially controlled by the electric field \cite{Ibrahim2016}, could be a key property in the design and fabrication of new types of spintronic and superconducting spintronic devices with multilevel characteristics. Probing similar magnetization spin reorientation experiments on unpatterned samples requires a magnetometer with a vector magnet or with 3D rotational capabilities. We expect such experiments to be carried out in a near future.

\section{Methods}

\subsection{Sample growth}

The basic system under investigation, MgO/Fe(10 nm)/MgO, represents a 10 nm thick Fe(100) soft layer interfaced by two 2 nm thick MgO(100) layers. The Fe layer between the two MgO barriers is continuous, as checked by in-situ RHEED and ex-situ AFM experiments. Moreover, after the growth of the first barrier it has been annealed for atomic-level flattening, as proved by in-situ RHEED. This insures model quality MgO/Fe/MgO interfaces. The MTJs were patterned by UV photolithography and Ar etching to an area of $20\times20\mu\text{m}^2$. The magnetic state of the Fe layer has been probed through TMR measurements, as it is interfaced via two MgO barriers by a magnetically hard (10 nm thick Fe and 20 nm thick Co) layer on one side and by a 40 nm thick normal metal Vanadium layer on the other. The full layer sequence is
V(40 nm)/MgO(2 nm)/Fe(10 nm)/MgO(2 nm)/Fe(10 nm)/Co(20 nm), represented in Figure \ref{esquema}d. All layers were deposited by molecular beam epitaxy at room temperature and ultra high vacuum conditions ($10^{-11} \text{ mbar}$). Details of the sample growth can be found in Ref.\cite{Tiusan2007}. 
The DC resistive measurements have been carried out at low bias (5 mV) at temperatures down to 5 K (i.e. above the superconducting critical temperature of Vanadium) using a JANIS He$^3$ cryostat equipped with a home-made superconducting vector magnet. The room temperature DC resistive measurements have been carried out in zero bias limit in a less shielded system which accounts for the apparently higher measurement noise in that case. Details of the main low temperature experimental setup were published previously \cite{Cascales2012}.

\subsection{Simulation methods}

The magnetic field dependence of the magnetization has been simulated at T=0 K by using MuMax3 code \cite{mumax}. The left part of the sketch shown in Figure \ref{esquema}d zooms the simulated free soft Fe layer interfaced by two MgO barriers. The right part in Figure \ref{esquema}d  shows both soft (Fe) and hard (Fe/Co) ferromagnets with the last one being fixed in-plane. The parameters used for Fe are: saturation magnetization $M_{sat}=1700 \times 10^{3} \, A/m$, exchange stiffness $A_{exch}=21 \times 10^{-12} \, \text{J/m}$, damping $\alpha=0.02$ and cubic anisotropy $K_c=4.8 \times 10^{4} \, \text{J/m}^3$. Interfacial layers with PMA include the surface anisotropy first order term $K_{s1}$ and (separately discussed) a second order term $K_{s2}$ anisotropy. The results have been confirmed to be independent on whether the MgO cells were represented as vacuum or as a weakly ($10^{-7}$) diamagnetic material. The parameters used for Cobalt are: $M_{sat}=1400 \times 10^{3}$A/m,  $A_{exch}=30 \times 10^{-12} \, \text{J/m}$ and $\alpha=0.02$. To approach the extensive in-plane dimension (in comparison with the thickness), the in-plane size of the simulated sample is set to $50\times50 \text{ nm}^2$ with periodic boundary conditions. The space was discretized in $16 \times 16 \times 147$ cells. This discretization has been chosen because it allows us to introduce in the micromagnetic problem the different models of interface anisotropies.

%%%%%%%%%%%%%%%%%%%%%%%%%%%%%%%%%%

\section{Acknowledgments}

The authors acknowledge NVIDIA academic GPU grant program, Raul Villar for critical reading of the manuscript, Jaroslav Fabian and Igor Zutic for discussions on the possible origin of the out of plane Rashba field and Pablo Andr\'es, Manuel Martin and Sara Garc\'ia for their help with measurements and simulations at the initial stages. This work has been supported in part by Spanish MINECO (MAT2015-66000-P, EUIN2017-87474), and Comunidad de Madrid (NANOFRONTMAG-CM S2013/MIT-2850). C.T. acknowledges "EMERSPIN" grant ID PN-III-P4-ID-PCE-2016-0143, No. UEFISCDI:22/12.07.2017.

\section{Author Contributions}

F.G.A. supervised the project and performed numerical simulations. I.M. performed experimental measurements and numerical simulations. The samples growth and photolithography were carried out by C.T and M.H., M.C. proposed the model explaining symmetry breaking. The paper was written by F.G.A. and I.M. with contributions and comments from all co-authors.

\section{Additional information}

\textbf{Competing Interests:} The authors declare that they have no competing interests.

\end{document}